\documentclass[runningheads]{llncs}

\usepackage{graphicx}
\usepackage{todonotes}
\usepackage{csquotes}
\usepackage{xcolor}
\usepackage{acro}

\DeclareAcronym{bdk}{
    short = {bdk},
    long = {Bitcoin Dev Kit},
    cite = {bdk}
}
\DeclareAcronym{HTLC}{
    short = {HTLC},
    long = {Hash Time-Locked Contract},
    short-indefinite = {an},
    long-indefinite = {a}
}
\DeclareAcronym{UTXO}{
    short = {UTXO},
    long = {Unspent Transaction Output}
}
\DeclareAcronym{CPFP}{
    short = {CPFP},
    long = {Child-pays-for-parent},
    cite = {cpfp}
}
\DeclareAcronym{p2p}{
    short = {p2p},
    long = {peer-to-peer}
}
\DeclareAcronym{API}{
    short = {API},
    long = {Application Programming Interface}
}
\DeclareAcronym{LN}{
    short = {LN},
    long = {Lightning Network},
    cite = {lightning2016}
}

\begin{document}

\title{Open problems in cross-chain protocols}

\author{
Thomas Eizinger\inst{1} \and
Philipp Hoenisch\inst{1} \and
Lucas Soriano del Pino\inst{1}
}
\institute{CoBloX Pty Ltd, Australia \email{\{firstname\}@coblox.tech}}

\maketitle

\begin{abstract}
Blockchain interoperability is a prominent research field which aims to build bridges between otherwise isolated blockchains.
With advances in cryptography, novel protocols are published by academia and applied in different applications and products in the industry.
In theory, these innovative protocols provide strong privacy and security guarantees by including formal proofs.
However, pure theoretical work often lacks the perspective of real world applications. 
In this work, we describe a number of hardly researched problems which developers encounter when building cross-chain products.

\end{abstract}

\section*{Introduction}

The domain of blockchain technology has been a prominent research field for industry and academia ever since Bitcoin was introduced in 2008 \cite{nakamoto2008}. 
Its central idea is simple: to provide a trustless and censorship-resistant way of transferring asset ownership between parties. 
Besides Bitcoin, a blockchain ecosystem has evolved over the years with hundreds of different implementations. Most blockchains provide their own coin which is used to pay for transaction fees, smart contract executions or as \textit{digital cash}. 

The age-old problem of interoperability, well studied in various other computer systems, is now also an important issue to be tackled in the ever evolving world of blockchain. 
In particular, \textit{cross-chain protocols} are an active area of research that tries to provide certain guarantees when moving coins between two blockchains.
Several protocols have been proposed over the past few years, e.g. hash-based locks, Scriptless Scripts and other more complex signature-based protocols.
Most of these proposals (rightfully) focus on the security aspects of the cryptography involved in the protocols, by including formal proofs or simply showing that they work \cite{comit2019}.

However, there is a whole new dimension of hardly studied and completely unsolved problems, which lies in between theoretical work and practical cross-chain product development.
In this work, we describe several of these with the intent of motivating anyone working in this field to collaborate on possible solutions.
This list is by no means exhaustive.
It instead represents the issues that we found to be most pressing in developing software for cross-chain protocols.

\section{Wallets}

Implementing cryptographic protocols on top of blockchains such as Bitcoin requires precise control over how the individual transactions are built.
For example, to unlock the funds within \iac{HTLC} on Bitcoin, one has to correctly construct the witness stack according to the semantics of the initial locking script.
In practice, this means having control of the private key within the software that knows about the protocol semantics, in order to produce the correct signatures.
This is an issue from a security perspective as a user has to share their private key with the software. 

Miniscript \cite{miniscript} represents an effort in trying to generalize how such spending conditions can be expressed.
A general way of expressing spending conditions allows existing wallets to support signing of arbitrary \acp{UTXO} as long as their script follows the miniscript language. 
Miniscript enables application to create complex spending conditions such as \acp{HTLC} and have transactions be signed by the user's wallet without the user having to share their private key with the application. 

Unfortunately, support for miniscript within wallets in the wild is basically non-existent.
Additionally, solutions like miniscript only work for spending conditions that are expressed as actual scripts.
Modern locking mechanisms such as adaptor signatures cannot be expressed using miniscript.

\subsection{Hot wallets}

Any software that implements cryptographic protocols currently has to roll its own hot wallet tailored to the needs of the protocol.
Efforts like \ac{bdk} attempt to make it easier to build such a wallet although it is still a non-trivial task.
In summary, we see two classes of problems here:

\begin{enumerate}
\item Lack of general solutions for expressing arbitrary locking mechanisms.
    
These would enable the development of \enquote{off-the-shelf} components that can be used to implement the required wallet functionalities for blockchain protocols without prior knowledge of specific protocols.
Consequently, it would be simpler to develop software implementing blockchain protocols and the user experience would improve by allowing users to utilize their existing wallets with the software providing the protocol implementation.

\item Lack of reusable components to safely roll your own wallet.

Reusable components to roll your own wallet would improve the ecosystem by generally accelerating development and allowing developers to focus on their protocol instead of having to implement a wallet from scratch.
\end{enumerate}

\subsection{Multi-currency wallets}

So far, we have looked at the problem space of wallets when implementing protocols on a single blockchain.
For cross-chain protocols, the problem space grows linearly with the number of chains supported by the protocol.
Different blockchains may use different elliptic curves and signature schemes, making it hard if not impossible to share algorithms or implementations between them.

\section{Blockchain monitoring and interaction}

A software implementing a cross-chain protocol is effectively a state machine that gets advanced by certain events happening on any of the blockchains involved.
Examples of such events are:

\begin{itemize}
    \item Blockchain time exceeds a certain time
    \item A specific UTXO is being spent
    \item A transaction reaches a certain confirmation target
\end{itemize}

To react to these events promptly, the software needs to be aware of the latest blockchain state at all times.
This is what we call \enquote{blockchain monitoring}.
There are several ways how an application can get access to the blockchain state:

\begin{enumerate}
    \item Talking to a self-hosted full node
    \item Talking to a shared, hosted full node
    \item Talking to a blockchain explorer
    \item Talking directly to other full nodes via the \ac{p2p} protocol
\end{enumerate}

We define the following three desired properties of blockchain monitoring:

\begin{itemize}
    \item Allow easy and fast on-boarding of the user
    \item Trustless and privacy preserving
    \item Efficient
\end{itemize}

In the following sections, we will go through the four possible monitoring options and show that none of them embody all three requirements.

\subsection{Self-hosted full node}

Running a full-node yourself provides by far the most flexibility for protocol software.
It allows for a simple poll-based model of querying for the latest blockchain state.
Given the self-hosted nature of the full node, such an implementation is reasonably efficient.
Additionally, accessing the network through a dedicated full node preserves the user's privacy and does not demand trusting a third-party.
Unfortunately, setting up full nodes is a time-intensive task due to the initial block download.
Furthermore, the resources required to continuously run multiple nodes - one per blockchain - are not to be underestimated.

\subsection{Shared, hosted full-node}

A shared, hosted full node allows for an easier setup compared to running one yourself.
However, privacy might be impacted and the user has to trust the node operator to provide an accurate view of the blockchain.
Finally, a node operator might charge the user based on the bandwidth used which has implications on how the software interacts with the node, i.e. the software cannot simply poll for state updates every 10 seconds.

\subsection{Blockchain explorer}

Accessing the blockchain via a block explorer enables easy and fast on-boarding of the user because it requires no prior setup.
Similar to a shared, hosted full-node, privacy might be impacted and trust in the operator of the blockchain explorer is required.
Blockchain explorers usually offer more high-level \acp{API} to interact with the blockchain.
For example, instead of having to query individual blocks and transactions, explorers usually already index the balances of addresses.
Such high-level \acp{API} drastically reduce the required network communication between the software and the explorer, thereby providing a reasonably efficient way of blockchain monitoring.

\subsection{\acs{p2p} protocol} % don't expand acronym in section title

Talking directly to other full nodes over the \ac{p2p} protocol of a blockchain network has several advantages over the previous options.
For one, it is as privacy preserving as running your own full node.
Second, depending on the capabilities of the \ac{p2p} protocol, it can be trustless.
For example, BIP157~\cite{bip157}{} allows for a client-side filtering of blocks.
Finally, accessing blockchain state directly via the \ac{p2p} protocol can also be more efficient than the other solutions due to decreased communication overhead and indirection.
Unfortunately, directly talking to other full nodes requires the application to apply all consensus rules itself to make sure it has a correct view of the latest state.
This turns out to be a non-trivial undertaking.
For example, for many blockchains, there is no clear specification of all consensus rules making them practically impossible to re-implement.

\subsection{Summary}

For a software implementing a cross-chain protocol, monitoring the blockchain is vital.
Without access to the latest state, the protocol's state machine cannot advance.
As the above sections show, efficient and trustless monitoring is currently only possible by running a full-node yourself.
That, in turn, greatly hinders the on-boarding process of users and is resource-intensive to operate.

\section{Fees}

Users have to pay transaction fees to get their transaction included in a block in almost any public blockchain system.
Whilst being useful to prevent spam and other attacks, they also present a problem in blockchain protocols that involve timelocks and are generally composed of more than a single transaction.

\subsection{Timelocks}
% \todo{\lucas{Maybe it's obvious but I don't think we've mentioned why timelocks appear in the fees section}, \thomas{The idea was that timelocks imply that you can predict the load and hence choose the fee correctly. Hence, protocols that use timelocks should also include, how to do that and not leave it up to the software :)}}
Timelocks are a common component of blockchain protocols.
They make it possible to express spending conditions that are not only based on knowledge of secrets like pre-images but based on time.
This is useful in allowing parties to abort or bail out of a protocol execution.

% Popular blockchains like Bitcoin have two kinds of timelocks.
% Absolute timelocks define a concrete point in time whereas relative timelocks are \textit{relative} to a certain event such as the inclusion of another transaction. 
% For cross-chain protocols, it is often desirable to rely on absolute timelocks to avoid attacks that speed up or slow down block generation on one of the participating chains.

% \todo{\thomas{in any case, timelocks are tricky because the load on the network cannot be reliably predicted, which means the fees a user needs to pay might differ from what they expected when the parameters of a protocol were negotiated (if enough time passes between those two events)}}

\subsection{Pre-signed transaction}

A pre-signed transaction is a transaction that is signed by one or more parties ahead of the time it is meant to be broadcasted.
Many blockchain protocols pre-sign punishment or refund transactions to provide safety to the parties involved.
For example, when setting up a \ac{LN} channel, the transaction to close the channel is signed before the transaction that actually opens it.

The mining fee that is paid by a transaction cannot be changed once it has been signed.
As such, pre-signing a transaction requires making an estimate of how large the mining fee should be.
Depending on the timeline over which the protocol operates, a forecast of the blockchain load can be extremely unreliable.
A varying network load can negatively impact the user if such a pre-signed transaction will no longer be confirmed within the block target that was originally anticipated.
One solution to this problem is the fee pumping technique \ac{CPFP} although similar to other topics mentioned in this work, such a strategy needs to be specifically developed and included in the software.

\subsection{Summary}

Timelocks as well as pre-signed transactions share a common characteristic:
Both concepts are about transactions that are to be broadcasted some time in the future instead of right away.
Given the dynamic nature of a blockchain's fee market, it is a non-trivial undertaking to reliably get a transaction confirmed within a certain block target \textit{starting from some point in the future}.

Cross-chain protocols often rely on both concepts, timelocks and pre-signed transactions, to achieve certain properties like atomicity or fairness.
Yet, in our experience, it is usually left up to the software and/or the user to ensure a certain transaction is included in the blockchain.
We see a lack of \enquote{off-the-shelf} solutions and general algorithms for reliably determining fees of such delayed transactions.

\section{Testing}

Like every software, implementations of cross-chain protocols need to be tested.
The ideal test suite combines a fast execution speed with a high degree of confidence that the software works as expected.
This combination allows for short feedback cycles and therefore greater productivity.
We have found that it is currently not possible to develop test suites for the cross-chain protocol space that fulfill both of these criteria.

For a transaction to be accepted by the network, a number of conditions must be met.
Some examples are:

\begin{itemize}
    \item The transaction must not spent an already spent output
    \item The signatures on the transaction must verify
    \item The transaction must not spend more coins than it consumes
    \item Any smart contract involved in the transaction must execute without errors
\end{itemize}

Some of these aspects can be verified in isolation.
For example, checking the validity of a signature is reasonably easy.
It follows that testing code that produces such a signature is also easy.

Verifying that a transaction does not double-spend an output is more complicated.
By nature, such a verification depends on the current state of the network.
Similarly, evaluating a smart contract also requires access to the current state.
Dependence on this state implies a more complicated test setup in which this state is constructed.

In the current state of affairs, creating such a state reliably is only possible by spinning up instances of the respective blockchain nodes with a \enquote{test} configuration.
Unfortunately, this drastically slows down the execution speed due to the start-up time and network communication overhead.
We've also found that running such tests in parallel can cause problems if all tests share the same nodes.
We experienced sporadic test failures caused by timeouts that don't happen with decreased parallelization.
On the other hand, spinning up a node per test to achieve isolation demands even more resources.

While testing against actual blockchain nodes provides a high level of confidence, these test suites tend to be slow and often need to be executed sequentially.
Attempts to run them in parallel easily lead to instability which lowers the confidence in the test suite.

In a cross-chain setting, the situation is even worse due to an increased number of combinations that need to be tested.

\section{Economics of atomic swaps}

Atomic swaps are a very popular cross-chain protocol.
Their name suggests that they represent an atomic \textit{swap} operation.
However, that is not quite true.
Atomic swaps merely present a time window within which both parties are committed to the swap.
Outside of this time window, no atomicity is guaranteed.

This has consequences for applications built on top of atomic swaps once the implementation details of this atomicity leak into the application layer.

\subsection{Draining attack}

Alice - by convention the party that moves first - can suffer from a draining attack where she sends the first transaction to the network without Bob ever moving forward afterwards.
Sending a transaction to the network requires Alice to pay mining fees yet Bob has no obligation to lock in his part of the swap.
Bob's inaction forces Alice to spend her coins via the \enquote{refund} path after a certain timeout has been reached, incurring in further fee expense.

\subsection{American Option}
If Bob decides to lock in his part of the swap, Alice is presented with an American Option to either take Bob's money - effectively executing the swap - or wait for the timeout, forcing both parties to spend their coins via the \enquote{refund} path.
Why would Alice make use of this option?
Once both parties have locked in their share, the rate is fixed.
Alice can now check how the price evolves on other markets and take the according action that is in her favor.

\subsection{Where is the atomicity?}

An atomic swap is atomic as long as both parties are committed to the swap.
In that case, it is a safe option to swap assets because no party has to make the full transfer first.

The problem here is the cost associated with achieving this atomicity:
Each party has to pay for a single transaction before they can actually execute the swap.

The asymmetry of the protocol leaks up into the application layer as you try to use atomic swaps to implement, for example, a trading platform.
Whoever moves first will be exposed to the draining problem.
Whoever redeems first holds an American Option.
Either problem is not pleasant to deal with as a market maker or service provider.

In summary, we want to point out that it is important to consider such kinds of scenarios in the design of cross-chain protocols.
To be easily and practically applicable, a protocol must not only be cryptographically sound, it must also present a clean abstraction that does not leak into the application layer.

%\section{Uncategorized}

%\subsection{Cross-asset vs cross-chain}

% Ethereum has demonstrated an high demand for trading mutliple assets within a single platform

%\subsection{Same-asset cross-chain protocols}

% i.e. move your USDT from Ethereum to Liquid

\section{Summary}

This work presents a number of problems that we consider open in the cross-chain protocol space from an industry perspective.
In several cases, the problems can be solved ad-hoc by, for example, implementing your own wallet with key management, blockchain monitoring, fee management and testing environments.
Having to deal with these problems slows down development if the actual goal is to implement a product on top of a cryptographic protocol.

% From the perspective of a Wardley Map\cite{wardly}, most of these problems reside in the \enquote{custom} column as there are no existing products of commodities that can be used. \phil{I'm not convinced that we should introduce Wardley here.}
% To drive the innovation forward, we need algorithms and protocols that solve problems like securely exposing arbitrary signing capabilities of wallets or blockchain monitoring in a general way such that reusable components and products can be built on top of them.

\bibliographystyle{splncs04}
\bibliography{samplepaper}

\begin{thebibliography}{1}
\providecommand{\url}[1]{\texttt{#1}}
\providecommand{\urlprefix}{URL }
\providecommand{\doi}[1]{https://doi.org/#1}

\bibitem{cpfp}
{Bitcoin Optech}: Child-pays-for-parent.
  \url{https://bitcoinops.org/en/topics/cpfp/} (2020), accessed: 2021-01-29

\bibitem{miniscript}
Blockstream: Miniscript - website. \url{http://bitcoin.sipa.be/miniscript/}
  (2020), accessed: 2021-01-25

\bibitem{comit2019}
COMIT: Connect all the blockchains!!!
  \url{https://medium.com/coblox/connect-all-the-blockchains-atomic-swap-78b38fff42e}
  (2018), accessed: 2021-01-13

\bibitem{bdk}
Filini, A., Casatta, R.: Bitcoin dev kit.
  \url{https://github.com/bitcoindevkit/bdk} (2020), accessed: 2021-01-29

\bibitem{nakamoto2008}
Nakamoto, S.: Bitcoin: A peer-to-peer electronic cash system.
  \url{https://bitcoin.org/bitcoin.pdf} (2008), accessed: 2021-01-13

\bibitem{bip157}
Osuntokun, O., Akselrod, A., Posen, J.: Client side block filtering.
  \url{https://github.com/bitcoin/bips/blob/master/bip-0157.mediawiki} (2017),
  accessed: 2021-01-29

\bibitem{lightning2016}
Poon, J., Dryja, T.: The bitcoin lightning network: Scalable off-chain instant
  payments (2016)

\end{thebibliography}
\end{document}